\documentclass{llncs}
\usepackage{amsfonts}
\usepackage{amsmath}
\usepackage{todonotes} 
\usepackage[utf8]{inputenc}

\usepackage{hyperref}

\DeclareMathOperator{\NDCG}{NDCG}
\DeclareMathOperator{\DCG}{DCG}
\DeclareMathOperator{\IDCG}{IDCG}

\DeclareMathOperator{\disc}{disc}

\DeclareMathOperator*{\argmin}{arg\,min}
\DeclareMathOperator*{\Var}{Var}

\begin{document}

\title{Enhancing LambdaMART Using Oblivious Trees}

\author{Marek Modrý\inst{1} \and Michal Ferov\inst{2}}

\institute{Seznam.cz, Radlická 3294/10, 150 00 Praha 5, Czech Republic\\
\email{marek.modry@firma.seznam.cz}
\and
Seznam.cz, Londýnské náměstí 856/2, 639 00 Brno, Czech Republic\\ 
\email{michal.ferov@firma.seznam.cz}}

\maketitle              
\begin{abstract}
Learning to rank is a machine learning technique broadly used in many areas such as document retrieval, collaborative filtering or question answering. We present experimental results which suggest that the performance of the current state-of-the-art learning to rank algorithm LambdaMART, when used for document retrieval for search engines, can be improved if standard regression trees are replaced by oblivious trees. This paper provides a comparison of both variants and our results demonstrate that the use of oblivious trees can improve the performance by more than $2.2\%$. Additional experimental analysis of the influence of a number of features and of a size of the training set is also provided and confirms the desirability of properties of oblivious decision trees.

\keywords{Document Retrieval for Search Engines, LambdaMART, Learning to Rank, Oblivious Decision Trees.}
\end{abstract}

\section{Introduction}
\label{sec:introduction}

\emph{Learning to rank}~(LTR) is a machine learning technique broadly used in many areas, such as document retrieval in search engines, collaborative filtering in recommender systems or question answering~\cite{liMcRankLTRUsingMultipleClassificationAndGradientBoosting2007,Cao:2007:FromPairwiseApproachToListwiseApproach,hang2011short}. Generally speaking, LTR is a subject of interest for any system that needs to order intermediate or final results with respect to a given utility function \cite{Burges2005Ranknet}.

The focus of this paper is document retrieval in the sense of search engine results ranking, which is also reflected by the data on which the experiments were conducted. 
There are many algorithms that approach the task in different ways but LambdaMART is the state-of-the-art LTR algorithm. It has been recently revealed that it is being used as a part of Facebook's machine learning framework~\cite{fb:ref}. In contrast to the most of LTR algorithms, LambdaMART uses direct optimisation with respect to a given objective function. In this paper we consider \emph{Normalised Discounted Cumulative Gain}. On the other hand, LambdaMART is prone to overfitting when training complicated models consisting of hundreds of trees. 

This paper proposes a modification of the standard LambdaMART algorithm which utilises oblivious decision trees instead of standard regression trees in the process of learning. Oblivious tree is a special kind of a decision tree with a constraint on the selection of decision rules. LambdaMART iteratively builds a model by constructing a sequence of decision trees whose predictions are summed up to obtain the final prediction. The introduction of oblivious trees brings many benefits like making the algorithm less prone to overfitting, simplifying and speeding up the use of the model and potentially decreasing the number of features included in the model. 

We present experimental results which suggest that the performance of the current state-of-the-art LTR algorithm LambdaMART can be improved if standard regression trees are replaced by oblivious trees. The experimental results were obtained both on a public LTR dataset and on a dataset originating from Czech search engine Seznam.cz. We performed the comparison using different parameter setups and also verified whether the size of the training set can influence the dominance of the oblivious trees variant.  

The paper is structured as follows. Existing literature and research is examined and described in Section~\ref{sec:relatedwork}. Formal definition of LTR framework is given in Section~\ref{sec:ltr} along with the description of performance evaluation for LTR models and description of LambdaMART algorithm and oblivious decision trees. Experiments, datasets and results are presented in Section~\ref{sec:experiments}. Finally, in Section~\ref{sec:conclusions} we provide summary of our results and suggestions for future work.

\section{Related Work}
\label{sec:relatedwork}
When solving standard classification and regression problems, the most common approach aims at direct optimisation of a given objective function. However, due to the non-differentiability of a performance measure function, direct optimisation is a challenging task in LTR. Hence, researchers approach LTR task in many different ways. 

For instance, LTR problem can be reformulated as a regression task of the relevance label prediction which approaches the problem in a pointwise manner. Each query-document pair is then considered a single data sample and the relations between documents belonging to a particular query are not taken into account. \emph{Mean Squared Error}~(MSE) is then usually used as the objective function~\cite{Breiman:2001:RandomForests}. Random Forest~\cite{Breiman:2001:RandomForests} or Multiple Additive Regression Trees~(MART)~\cite{Friedman:2000:GreedyFunctionApproximationAGradientBoostingMachine} can be utilised to solve the task in the aforementioned manner. Similarly, PRank algorithm proposed in \cite{Crammer:2001:PranikingWithRanking} uses a neural network to predict the relevance label. However, the authors of~\cite{Crammer:2001:PranikingWithRanking} extend the task to ordinal regression, where the relevance score is converted to the relevance class~(resp. label) in the end. Besides, there are also pointwise algorithms that treat the problem as a classification problem.
For instance, McRank algorithm~\cite{liMcRankLTRUsingMultipleClassificationAndGradientBoosting2007} uses gradient boosting tree algorithm and reformulates the task as a multiple ordinal classification.

Algorithms applying a pairwise approach formalise the problem as classification or regression on pairs of query-documents. As was pointed out in~\cite{Cao:2007:FromPairwiseApproachToListwiseApproach}, even though pairwise formalisations benefit from the possibility of using existing classification or regression methods, the results can be suboptimal as the models optimise surrogate loss functions, the computation efficiency can be a problem and the results can be potentially biased towards queries with more documents. In~\cite{herbrichRankingSVM2000}, RankingSVM algorithm employs ordinal regression to determine relative relevance of document pairs. RankBoost~\cite{FreundRankBoost2003} is a boosting algorithm based on AdaBoost's idea and uses a sequence of weak learners in order to minimise the number of incorrectly ordered pairs. Burges et al.~\cite{Burges2005Ranknet} proposed RankNet algorithm that learns a neural network to predict the relevance score of a single query-document in such a way that the score can be used to correctly order any pair of query-document samples. The network is optimised using gradient descent on a probabilistic cost function defined on pairs of documents.

Algorithms taking the whole ranking list into account belong to the group of list-wise algorithms. The approach is straightforward and uses all information about the ranked list to further improve the model. On the other hand, direct optimisation is very challenging. Authors of PermuRank~\cite{XuLiudirectlyoptimizingPermurank2008} use SVM technique to minimise a hinge loss function on permutations of documents. Similarly, AdaRank~\cite{XuLiAdaRank2007} repeatedly constructs weak rankers in order to minimise an exponential loss which is derived from the original performance measure. Examples of other algorithms that employ list-wise approach are ListMLE~\cite{XiaListWiseApproachToLTRTheoryAndAlgorithm2008}, ListNet~\cite{Cao:2007:FromPairwiseApproachToListwiseApproach}, RankCosine~\cite{Qin:2008:QueryLevelLossFunctions}, LambdaRank~\cite{burgesLambdaRank2006} or LambdaMART~\cite{burges:2010:FromRanknetToLambdaRankToLambdaMART}. Note that LambdaRank was the first algorithm to propose using lambdas to define gradients. LambdaMART algorithm is the main focus of this paper.

\emph{Oblivious decision tree} is a special kind of a decision tree with constraints on the selection of a decision rule. Experimental results of Almuallim and Dietterich~\cite{almuallim1991learning} demonstrated that standard decision trees, e.g. those built using ID3 algorithm, can perform poorly on datasets with many irrelevant features. This problem is addressed by Langley and Sage in~\cite{langley1994oblivious} where they proposed tackling the problem of irrelevant features by using oblivious decision trees. The constraints on decision rules selection were introduced also by Schlimmer in~\cite{schlimmer1993efficiently}. Although our modification uses a basic greedy top-down induction of oblivious trees, there have been several methods of oblivious tree construction proposed (see \cite{rokach:2008:datamining,kohavi1995oblivious,last2002improving}). Authors of YetiRank algorithm~\cite{gulin2011winning} introduced oblivious trees into LTR task. However, YetiRank works in a different way and utilises oblivious trees differently than LambdaMART.

\section{Learning to Rank}
\label{sec:ltr}

\subsection{Formal description of LTR}
\label{sec:problemdesc}
LTR can be formally described as follows. In the training phase of LTR process, a set of queries $Q = \{q_1,q_2,\dots,q_n\}$, where $n$ denotes the number of queries, is given. For every query $q_i$ there is a set of documents $\mathbf{d_i} = \{d^i_1,d^i_2,\dots,d^i_{m(q_i)}\}$, where $m(q_i)$ is the number of documents given for the query $q_i$. For every query-document pair $(q_i, d^i_j)$, where $i \in \{1, \dots, n\}$ and $j \in\{1, \dots, m(q_i)\}$, a label $y^i_j$ is given. Similarly, a feature vector $\vec{x}^i_j \in \mathbb{R}^{l}$, where $l$ is the number of features, is specified for each query-document pair $(q_i,d^i_j)$, where $i \in \{1,2,\dots,n\}$ and $j \in \{1,\dots,m(q_i)\}$. Finally, the dataset $S$ can be defined as 
\begin{equation}
	\label{eq:traindataset}
		S = \left\{(q_i,\mathbf{d_i},\mathbf{y_i})\right\}^n_{i=1},
\end{equation}
where $\mathbf{y_i} = \{y^i_1, \dots, y^i_{m(q_i)}\}$ for $i \in \{1, \dots, n\}$.

In our formal framework, the model is a function $f$ mapping a feature vectors corresponding to samples to score values, i.e. $f: \mathbb{R}^l \to \mathbb{R}$. By evaluation of the model $f$ on the query-document sample $(q_i, d^i_j)$ we mean application of the function $f$ on the feature vector $\vec{x}^i_j$. When applying the model, all query-document pairs $(q_i,d^i_j)$ in the list corresponding to the query $q_i$ are evaluated and have their score assigned. The list is then ordered according to the score in the descending order.
 
A permutation $\pi_i(\mathbf{d_i},f)$ of the set of integers $\{1, \dots, m(q_i)\}$ is created based on the document list $\mathbf{d_i}$ and the model function $f$. To be more specific, the permutation represents a ranked (ordered) list and $\pi_i(j)$ denotes the position of the document $d^i_j$ in the ranking (ordering). When applying the model, $f$ is utilised to rank elements of documents lists for each single query in the dataset $S$ and, finally, the performance of the model is evaluated using given evaluation measure (e.g. NDCG).

\subsection{Performance evaluation}
Unlike the models built during classification or regression tasks, a quality of LTR model depends primarily on the order of a sorted list. As a result, LTR evaluation functions usually are not smooth or differentiable, which makes the direct optimisation very challenging as we cannot use gradient descent methods. In the following paragraphs, \emph{Normalised Discounted Cumulative Gain}~(NDCG) measure is introduced. Our experiments described in the text below are evaluated using NDCG. 
 
As described in \cite{Jarvelin:2002:CumulatedGainBasedEvaluationOfIRTechniques}, NDCG is applicable to multi-graded relevance scale. \emph{Discounted Cumulative Gain} (DCG) can be defined as follows:

\begin{equation}
\label{equation:dcg}
\DCG(f;\mathbf{d_j}, \mathbf{y_j}) = \sum_{i=1}^{m}G(y(d_{{\pi_f(i)}}))) \disc(i),
\end{equation}
where $\mathbf{d_j} = \{d_1, \dots, d_m\}$ is a set of query-document pairs belonging to query $q_j$ and $\mathbf{y_j} = \{y(d_1), \dots, y(d_m)\}$ is the set of corresponding relevance labels, $G \colon \mathbb{R} \to \mathbb{R}$ is an increasing \emph{gain function}, $\disc \colon \mathbb{R} \to \mathbb{R}$ is a decreasing \emph{position discount function}, $\pi_{f}$ is the resulting ranked list and $d_{{\pi_f(i)}}$ is the document at $i^{th}$ position in the ranked list. \emph{Ideal Discounted Cumulative Gain} ($\IDCG$) represents the DCG score of a list that was ranked in the best way possible:

\begin{equation}
\label{equation:idcg}
\IDCG(f;\mathbf{d_j}, \mathbf{y_j}) = \max_{\pi_f} \left\{ \sum_{i=1}^{m}G(y(d_{{\pi_f(i)}}))) \disc(i) \right\}
\end{equation}

Subsequently, NDCG is defined in a straightforward way:
\begin{equation}
\label{equation:ndcg}
\NDCG(f;\mathbf{d_j}, \mathbf{y_j}) = \frac{\DCG(f;\mathbf{d_j}, \mathbf{y_j})}{\IDCG(f;\mathbf{d_j}, \mathbf{y_j})}.
\end{equation}

Following ~\cite{Wu:2010:AdaptingBoostingForInformationRetrievalMeasures} and~\cite{liMcRankLTRUsingMultipleClassificationAndGradientBoosting2007}, the \emph{gain function} $G$ is defined as $G(r) = 2^r-1$ and the \emph{discount function} $\disc$ is defined as 
\begin{displaymath}
	\disc(i)= 
	\begin{cases}
		\frac{1}{\log_{2}(i+z)}	&\mbox{ if }i \le C,\\
		 0 						& \mbox{ if } i > C,
	\end{cases}
\end{displaymath}
where $C$ is a fixed integer \emph{cut-off constant} that specifies on how many top documents does the measure focus. The use of a cut-off constant can be motivated for instance by the paging of search engine results and limited number of initially displayed documents as noted in ~\cite{Burges2005Ranknet,Wu:2010:AdaptingBoostingForInformationRetrievalMeasures}. The usage of cut-off constant $k$ and limiting the evaluation only to the top $k$ documents is denoted by $\NDCG@k$. When calculating performance on multiple queries, the resulting performance is the mean of all performance scores averaged over the corresponding set of queries. 

\subsection{LambdaMART}
\label{sec:lambdamart}
LambdaMART is a LTR algorithm introduced in~\cite{Wu:2010:AdaptingBoostingForInformationRetrievalMeasures}. It combines regression trees boosting technique that is utilised in MART\footnote{Multiple Additive Regression Trees} algorithm ~\cite{Friedman:2000:GreedyFunctionApproximationAGradientBoostingMachine} and LambdaRank's idea of using lambda trick that avoids the non-smoothness problem of ranking performance measures~\cite{burgesLambdaRank2006}. 

There is no known way to derive the gradients from the performance measure directly. LambdaMART overcomes this problem by computing the gradients based on the current state of the model and desirable changes in the order of elements in the list, so called \emph{lambdas}. Lambda coefficient represents how the model's prediction for a particular document should change in the next iteration in order to improve the model performance. As LambdaMART builds one tree per iteration, lambdas representing the target function $t_{k+1} \colon \mathbb{R}^l \to R$ for the $(k+1)^{\text{th}}$ tree are computed after building $k$ trees. Note that the direct optimisation following the computed gradients (i.e. lambdas) can be further modified by using \emph{Newton's method} (see \cite{whittaker1967newton}). Newton's method is implemented by adjusting values in the leaves of a newly built tree. For detailed explanation of the Newton's method step in LambdaMART see~\cite{burges:2010:FromRanknetToLambdaRankToLambdaMART}. 

\subsection{Oblivious Decision Trees}
\label{sec:oblivioustrees}
Oblivious decision tree is a particular kind of a decision tree. It is characterised by the constraint which allows to select only one feature in a particular level of the decision tree, i.e. all the decision rules in the specific level can involve only one selected feature. Although the definition of oblivious decision trees in~\cite{rokach:2008:datamining} permits to use different threshold values for the selected feature, our implementation goes even further and all nodes in the given depth use only a single rule (i.e. feature and threshold is uniform on levels).

In \cite{kohavi:1998:targeting} Kohavi et al. proposed a way to represent oblivious decision trees by decision tables. A decision table can be implemented very efficiently and can be processed much faster than a standard decision tree with no constraints. Such efficiency advantage can be essential for commercial search engines (such as Seznam.cz) which have to process hundreds of requests per second.

Furthermore, as the authors of \cite{almuallim1994learning,maimon2002improving,schlimmer1993efficiently} have noted, the constraint of oblivious decision trees has a positive effect on the effectiveness of feature selection which is a desired property, especially when dealing with datasets containing many uninformative irrelevant features. 

Our implementation uses top-down greedy algorithm for oblivious trees induction. A decision rule is selected for each level of the tree in a top-down manner. Each rule consists of a feature and a threshold value which splits $k$ disjoint sets into $2k$ disjoint subsets. Note that as we are working with complete binary trees $k = 2^{k'}$ where $k'$ is the number of levels of the tree. Formally speaking, a splitting rule $r$ is a pair $(i, v)$, where $i \in \{1, \dots, l\}$ is a index of a feature and $v \in \mathbb{R}$ is a threshold value. A set of feature vectors $X \subseteq \mathbb{R}$ can be then split into two disjoint subsets $X^L, X^R$ such that for every $\vec{x} = (x_1, \dots, x_l) \in X$ we have $x \in X^L$ if $x_i \leq v$ and $x \in X^R$ if $x_i > v$. Naturally, splitting rule can be applied to any collection of $k$ disjoint sets $X_1, \dots, X_k \subseteq \mathbb{R}^l$ to obtain a collection of $2k$ disjoint sets $X_1^L, X_1^R, \dots, X_k^L, X_k^R$. When given a collection of sets $\mathcal{X} = \{X_1, \dots, X_k\} \subseteq \mathcal{P}(\mathbb{R}^l$) and a target function $t \colon \mathbb{R}^l \to \mathbb{R}$, the splitting rule is then chosen as
\begin{equation}
   \argmin_r\left\{M(r, t, \mathcal{X}) \right\},
\end{equation} 
where $M$ is a non-negative real-valued function measuring the optimality of the splitting rule $r$ on the collection $X_1, \dots, X_k$ with respect to the target function $t$ (with a slight abuse of notation we can assume that $M$ is defined for every $k \in \mathbb{N}$). In our implementation, the optimality measure $M$ was defined as 
\begin{equation}
\label{equation:splitcrit}
M(r, t, \mathcal{X}) = \frac{1}{\sum_{i=1}^k |X_i|} \left(\sum_{i=1}^{k} |X_i^L| \Var\left(t(X_i^L)\right) + |X_i^R| \Var\left(t(X_i^R)\right)\right),
\end{equation}
where $t(X_i) = \{t(\vec{x}) \mid \vec{x} \in X_i \subseteq \mathbb{R}^l\}$ is the set of target scores corresponding to the samples in the set (resp. leaf node) $X_i$.

\section{Experiments}
\label{sec:experiments}

\subsection{Data}
\label{sec:experiments:data}

The experiments were performed using two datasets. MSLR-WEB10k dataset\footnote{Microsoft Learning to Rank Datasets, 2010. Available at \url{http://research.microsoft.com/mslr}}, which is a public LTR dataset released in 2010 by Microsoft, and the second dataset originating from Czech search engine Seznam.cz. Statistics of both datasets are provided in Tab.~\ref{tb:datastats}.

\begin{table}[h]
\centering
\caption{Dataset statistics}
\label{tb:datastats}
\begin{tabular}{lcc|c|c|c|c|c}
\cline{4-7}
\textbf{}                       & \textbf{}                      & \textbf{} & \multicolumn{4}{l|}{\textbf{Documents per Query}} & \textbf{}                \\ \hline
\multicolumn{1}{|l|}{\textbf{Dataset}} & \multicolumn{1}{l|}{\textbf{Queries}} & \textbf{Lines} & \textbf{Mean} & \textbf{Median} & \textbf{Max} & \textbf{Min} & \multicolumn{1}{l|}{\textbf{Features}} \\ \hline
\multicolumn{1}{|l|}{\textbf{Seznam.cz data}} & \multicolumn{1}{c|}{25728} & 1103295 & 42.88 & 40 & 265 & 1 & \multicolumn{1}{c|}{84} \\ \hline
\multicolumn{1}{|l|}{\textbf{MSLR-WEB10K}} & \multicolumn{1}{c|}{10000} & 1200192 & 120.02 & 110 & 908 & 1 & \multicolumn{1}{c|}{136} \\ \hline
\end{tabular}
\end{table}

MSLR-WEB10k dataset was chosen due to its availability, its sufficient size and also its multi-graded relevance labels. On the other hand, as the authors of the dataset claim (see~\cite{mslr:ref}), features in MSLR-WEB10k dataset are \textit{"those widely used in the research community"}. Although Seznam.cz's dataset is not publicly accessible, it allowed us to carry out the experiments on real-life data with a subset of features that are used in a commercial search engine.

The datasets, similarly to the most of Learning to rank datasets, use LibSVM (also called SVM\textsuperscript{light}) format, i.e. each line represents a query-document pair, specifying its relevance label, query ID and features. Both datasets use multi-graded scale of relevance labels that were annotated by human assessors. The higher the label, the more relevant the document is to a given query. While MSLR-WEB10k uses 5 grades scale, Seznam.cz uses 6 grades. Both datasets were aligned to use $1$ as the minimal relevance label in the range, i.e. in MSLR-WEB10k dataset $y^i_j = 5$ means that the document $d_j$ is \emph{maximally relevant} to the query $q_i$ whereas $y^i_j = 1$ means that $d_j$ is \emph{completely irrelevant} to $q_i$.

Both datasets were split into 5 folds that were subsequently used for cross-validation. Training dataset, validation set and testing set therefore consist of $60\%$, $20\%$ and $20\%$ of queries, respectively. Note that each query with its documents represent a compact data sample which means that a single query cannot be divided into more folds and therefore the dataset is split query-wise.

\subsection{Methodology}
\label{sec:methodology}

In this paper, two LambdaMART algorithm variants are compared, one using standard regression trees (as described in~\cite{Wu:2010:AdaptingBoostingForInformationRetrievalMeasures}) and one using oblivious decision trees instead. While RankLib~\cite{ranklib:ref} library was used to train standard LambdaMART models, Seznam.cz's own implementation of LambdaMART with oblivious decision trees called \emph{RCRank} was used to train the other variant. 

As noted above, the comparison was performed using cross-validation with training, validation and test sets consisting of $60\%$, $20\%$ and $20\%$ queries, respectively. There were 5 runs of each experiment using the same parameter setup but different data folds. Validation dataset was used to find the optimal number of trees in order to avoid overfitting and final test set performance was obtained by averaging the scores over experiments with the same parameter setup. 

Since the optimal parameters for both LambdaMART variants can differ, experiments on various parameter setups for each of the algorithm variants were performed and subsequently compared. Total number of leaves in a tree and learning rate were subjects of the change in particular setups. With 1000 trees being the maximal size of the forest, the length of each forest was cut according to the validation set to the size with highest validation quality (measured by NDCG@10). Lambda gradients in both algorithm variants also optimise NDCG@10 measure.

The second set of experiments was focused on the influence of the size of the training set. Although the setup of parameters were the same as in the previous experiments, the size of the training set was decreased. Originally, the training set consisted of $60\%$ (i.e. approximately 700~000 document-query pairs) of all samples which was lowered to only $6\%$ of samples (i.e. approximately 70~000 query-document pairs) and the training set in a subsequent experiment consisted only of $0.85\%$. samples~(i.e. approximately 10~000 query-document pairs). Note that, apart from the size of the training set, nothing was changed with respect to the the first set of experiments. Therefore the parameters, just like the size and splitting strategy of validation set and test set, remained the same. This experiment was performed only on MSLR-WEB10k dataset.

Analysis of how the number of features can influence the performance of the algorithms was performed on a modified dataset (the dataset with 10 times smaller training set was used). The list of features was sorted by the number of occurrences in models in the first set of experiments and only 50 most frequent features were kept in the dataset. We aimed at removing such features that are irrelevant or noisy.

\subsection{Results}
\label{sec:results}
The first set of experiments aimed at comparing the two algorithm variants, LambdaMART~(using regression trees) and RCRank~(using oblivious trees). Tab.~\ref{table:comp:seznam} and Tab.~\ref{table:comp:mslr} presents the results of the comparison on both, public and internal dataset. LambdaMART does not outperform RCRank in no parameter setup. When comparing the best results achieved by each of the algorithms, RCRank reached NDCG@10 score of $0.7135$ and $0.5706$, while LambdaMART's score was $0.7110$ and $0.5582$, on Seznam.cz and MSLR-WEB10k datasets, respectively, which means improvement by $0.35\%$ and $2.22\%$. Despite the improvement on Seznam.cz dataset being slightly lower, the improvement is consistent over all parameter setups. 

\begin{table}[h]
\centering
\caption{Comparison of RCRank and LambdaMART on Seznam.cz Dataset using NDCG@10 and different settings of learning rate and leaves number. RC and LM stands for RCRank and LambdaMART. Bold font is used for the best results of each algorithms.}
\label{table:comp:seznam}
\begin{tabular}{c|c|c||c|c||c|c||c|c||c|c|}

\cline{2-11} & \multicolumn{10}{c|}{\textit{\textbf{learning rate}}}  \\ \hline
\multicolumn{1}{|c|}{\textit{\textbf{leaf}}}& \multicolumn{2}{c||}{\textbf{0.11}} & \multicolumn{2}{c||}{\textbf{0.13}} & \multicolumn{2}{c||}{\textbf{0.15}} & \multicolumn{2}{c||}{\textbf{0.17}} & \multicolumn{2}{c|}{\textbf{0.19}} \\ \hline
\multicolumn{1}{|c|}{\textbf{8}}& 0.7036 & 0.7015 & 0.7045 & 0.7018 & 0.7056 & 0.7023 & 0.7062 & 0.7020 & 0.7067 & 0.7026 \\ \hline
\multicolumn{1}{|c|}{\textbf{16}}& 0.7073 & 0.7056 & 0.7082 & 0.7062 & 0.7085 & 0.7049 & 0.7089 & 0.7063 & 0.7097 & 0.7057 \\ \hline
\multicolumn{1}{|c|}{\textbf{32}}& 0.7097 & 0.7080 & 0.7106 & 0.7086 & 0.7112 & 0.7080 & 0.7115 & 0.7073 & 0.7118 & 0.7077 \\ \hline
\multicolumn{1}{|c|}{\textbf{64}}& 0.7121 &  \textbf{0.7110} & 0.7128 &  \textbf{0.7110} & \textbf{0.7135} &  \textbf{0.7101} & \textbf{0.7130} & 0.7094 & \textbf{0.7131} & 0.7096 \\ \hline
 & \textit{RC}      & \textit{LM}     & \textit{RC}      & \textit{LM}     & \textit{RC}      & \textit{LM}     & \textit{RC}      & \textit{LM} & \textit{RC}      & \textit{LM}     \\ \cline{2-11}
\end{tabular}
\end{table}

\begin{table}[h]
\centering
\caption{Comparison of RCRank and LambdaMART on MSLR-WEB10k using NDCG@10 and different settings of learning rate and leaves number. Bold font is used for the best results of each algorithms.}
\label{table:comp:mslr}
\begin{tabular}{c|c|c||c|c||c|c||c|c||c|c|}

\cline{2-11} & \multicolumn{10}{c|}{\textit{\textbf{learning rate}}}  \\ \hline
\multicolumn{1}{|c|}{\textit{\textbf{leaf}}}& \multicolumn{2}{c||}{\textbf{0.11}} & \multicolumn{2}{c||}{\textbf{0.13}} & \multicolumn{2}{c||}{\textbf{0.15}} & \multicolumn{2}{c||}{\textbf{0.17}} & \multicolumn{2}{c|}{\textbf{0.19}} \\ \hline
\multicolumn{1}{|c|}{\textbf{8}}& 0.5642 & 0.5529 & 0.5652 & 0.5528 & 0.5656 & 0.5540 & 0.5663 & 0.5536 & 0.5668 & 0.5546 \\ \hline
\multicolumn{1}{|c|}{\textbf{16}}& 0.5671 & 0.5564 & 0.5684 & 0.5554 & 0.5686 & 0.5557 & 0.5680 & 0.5557 & 0.5684 & 0.5547 \\ \hline
\multicolumn{1}{|c|}{\textbf{32}}& \textbf{0.5701} & 0.5572 & 0.5695 & 0.5577 & 0.5696 & \textbf{0.5570} & 0.5696 & 0.5567 & 0.5682 & 0.5571 \\ \hline
\multicolumn{1}{|c|}{\textbf{64}}& \textbf{0.5706} & \textbf{0.5582} & \textbf{0.5706} & \textbf{0.5580} & 0.5701 & 0.5575 & 0.5688 & 0.5571 & 0.5676 & 0.5561 \\ \hline
 & \textit{RC}      & \textit{LM}     & \textit{RC}      & \textit{LM}     & \textit{RC}      & \textit{LM}     & \textit{RC}      & \textit{LM} & \textit{RC}      & \textit{LM}     \\ \cline{2-11}
\end{tabular}
\end{table}

In order to analyse the influence of the size of the training set on RCRank's superiority, two experiments set on smaller training sets were performed. Tab.~\ref{table:comp:mslr:small} presents results of models built using training set ten times smaller. It can be observed that RCRank outperforms LambdaMART even when much smaller data set is being used. While NDCG score of RCRank models decreased by $5.12\%$ on average, the scores of LambdaMART models decreased by $5.64\%$ which could be a sign of oblivious trees being less prone to overfitting on smaller datasets.

\begin{table}[h]
\centering
\caption{Comparison of RCRank and LambdaMART on MSLR-WEB10k with 10~times smaller training set using NDCG@10 and different parameter settings. Bold font is used for the best results of each algorithms. }
\label{table:comp:mslr:small}
\begin{tabular}{c|c|c||c|c||c|c||c|c||c|c|}

\cline{2-11} & \multicolumn{10}{c|}{\textit{\textbf{learning rate}}}  \\ \hline
\multicolumn{1}{|l|}{\textit{\textbf{leaf}}}& \multicolumn{2}{c||}{\textbf{0.11}} & \multicolumn{2}{c||}{\textbf{0.13}} & \multicolumn{2}{c||}{\textbf{0.15}} & \multicolumn{2}{c||}{\textbf{0.17}} & \multicolumn{2}{c|}{\textbf{0.19}} \\ \hline
\multicolumn{1}{|c|}{\textbf{8}}& \textbf{0.5410} & 0.5241 & 0.5401 & 0.5243 & 0.5400 & 0.5251 & 0.5376 & 0.5238 & 0.5377 & 0.5233 \\ \hline
\multicolumn{1}{|c|}{\textbf{16}}& 0.5408 & \textbf{0.5268} & 0.5383 & \textbf{0.5263} & 0.5388 & 0.5256 & 0.5380 & 0.5249 & 0.5380 & 0.5246 \\ \hline
\multicolumn{1}{|c|}{\textbf{32}}& \textbf{0.5421} & \textbf{0.5266} & 0.5395 & 0.5250 & 0.5393 & 0.5253 & 0.5390 & 0.5244 & 0.5372 & 0.5242 \\ \hline
\multicolumn{1}{|c|}{\textbf{64}}& \textbf{0.5412} & 0.5259 & 0.5403 & 0.5245 & 0.5385 & 0.5233 & 0.5379 & 0.5212 & 0.5364 & 0.5221 \\ \hline
 & \textit{RC}      & \textit{LM}     & \textit{RC}      & \textit{LM}     & \textit{RC}      & \textit{LM}     & \textit{RC}      & \textit{LM} & \textit{RC}      & \textit{LM}    \\ \cline{2-11}
\end{tabular}
\end{table}

Similar results were achieved in experiment using 70~times less samples. The score of RCRank decreased by $10.44\%$ on average and the score of LambdaMART models decreased by $11.15\%$ in comparison to the full-size dataset. Best models' NDCG@10 scores for RCRank and LambdaMART were $0.5115$ and $0.4988$.

\begin{table}[ht]
\centering
\caption{Comparison of RCRank and LambdaMART on MSLR-WEB10k with 10 times smaller training set and only 50 selected features using NDCG@10 and different parameter settings. Bold font is used for the best results of each algorithms. }
\label{table:comp:mslr:features}
\begin{tabular}{c|c|c||c|c||c|c||c|c||c|c|}
\cline{2-11} & \multicolumn{10}{c|}{\textit{\textbf{learning rate}}}  \\ \hline                                                                                                                                      
\multicolumn{1}{|l|}{\textit{\textbf{leaf}}}& \multicolumn{2}{c||}{\textbf{0.11}} & \multicolumn{2}{c||}{\textbf{0.13}} & \multicolumn{2}{c||}{\textbf{0.15}} & \multicolumn{2}{c||}{\textbf{0.17}} & \multicolumn{2}{c||}{\textbf{0.19}} \\ \hline                                                                                       
\multicolumn{1}{|c|}{\textbf{8}}& 0.5400 & 0.5263 & 0.5405 & 0.5269 & 0.5395 & 0.5265 & 0.5401 & 0.5268 & 0.5399 & 0.5254 \\ \hline 
\multicolumn{1}{|c|}{\textbf{16}}& \textbf{0.5422} & \textbf{0.5292} & 0.5412 & \textbf{0.5288} & 0.5402 & \textbf{0.5286} & 0.5401 & 0.5279 & 0.5385 & 0.5273 \\ \hline 
\multicolumn{1}{|c|}{\textbf{32}}& 0.\textbf{5417} & 0.5278 & 0.5403 & 0.5272 & 0.5399 & 0.5270 & 0.5384 & 0.5255 & 0.5379 & 0.5271 \\ \hline 
\multicolumn{1}{|c|}{\textbf{64}}& 0.\textbf{5413} & 0.5268 & 0.5390 & 0.5255 & 0.5368 & 0.5248 & 0.5376 & 0.5222 & 0.5363 & 0.5227 \\ \hline  
 & \textit{RC}      & \textit{LM}     & \textit{RC}      & \textit{LM}     & \textit{RC}      & \textit{LM}     & \textit{RC}      & \textit{LM} & \textit{RC}      & \textit{LM}    \\ \cline{2-11}
 
\end{tabular}
\end{table}

The last set of experiments analysed the influence of number of features. Our hypothesis was that the removal of less important features will affect  RCRank and LambdaMART differently. As mentioned above, standard regression trees can perform worse when irrelevant features are present in the dataset, whereas oblivious trees should not be influenced by irrelevant features as much as standard regression trees. Tab.~\ref{table:comp:mslr:features} presents the results of the experiments performed on MSLR-WEB10k dataset with smaller training sets. 

Interestingly, even though 86 features were removed from the dataset, overall performance of LambdaMART improved with average improvement $0.37\%$ (averaged over all parameter setups). Since RCRank's overall performance did not improve for each of the parameters setup and the average score increased only by $0.09\%$, LambdaMART is more prone to be affected by the presence of less relevant features. Note that results of all comparison experiments are statistically significant (\emph{p-value} 0.01) except for a single one.\footnote{Experiment on trees with 64 leaves and learning rate 0.11 on Seznam.cz dataset.}

\section{Conclusions}
\label{sec:conclusions}
In this paper, we proposed an enhancement of LambdaMART algorithm substituting standard regression trees for oblivious decision trees which have several desirable properties. Using both, public MSLR-WEB10k dataset and internal dataset of Seznam.cz search engine, we experimentally compared performances of both variants. Moreover, the influence of various changes in the size of the dataset was examined. 

Depending on the dataset, we showed that by using oblivious trees instead of standard regression trees, the performance can be increased by more than $2.2\%$. Our experimental results further demonstrate that oblivious trees variant deals slightly better with smaller training datasets, as NDCG score of LambdaMART decreases faster than the score of RCRank\footnote{Our implementation of LambdaMART using oblivious trees} with the decrease of the size of training set. 

On the other hand, when potentially noisy and irrelevant features are removed from the dataset, LambdaMART's score improves more than the score of RCRank. This can be a sign confirming the aforementioned fact that standard regression trees can perform worse when irrelevant features are present in the training data.

Future work could focus on better theoretical understanding of the relation between standard regression trees and oblivious trees. Furthermore, all analysis could be also performed using various ranking measures, such as Expected Reciprocal Rank. Last but not least, it would be very useful to specify the properties of a dataset which is favourable to model using LambdaMART and not RCRank and vice versa.

\section*{Acknowledgements}
The authors would like to thank their colleagues from Seznam.cz research group for many insightful discussions, in particular to the authors of the original RCRank implementation that was based on Gradient Boosted Regression Trees without lambdas, Tomáš Cícha and Roman Rožník.

\bibliographystyle{abbrv}
\bibliography{references}
\end{document}